\documentclass{Interspeech}

\usepackage{amsmath}
\usepackage{amssymb}
\usepackage{mathtools}
\usepackage{amsthm}
\usepackage{algpseudocode}
\usepackage{algorithm}
\usepackage{etoolbox}
\usepackage{amsmath, amssymb, amsthm} 
\usepackage{mathrsfs}
\usepackage{etoolbox}
\usepackage{stmaryrd}
\usepackage{graphicx}
\usepackage{subcaption}

\usepackage[utf8]{inputenc} 
\usepackage[T1]{fontenc}    
\usepackage{hyperref}       
\usepackage{url}            
\usepackage{booktabs}       
\usepackage{amsfonts}       
\usepackage{nicefrac}       
\usepackage{xcolor}         

\usepackage[capitalize,noabbrev]{cleveref}
\usepackage[textsize=tiny]{todonotes}

\usepackage{bm}

\def\eqref#1{equation~\ref{#1}}

\def\1{\bm{1}}

\def\vc{{\bm{c}}}

\def\vp{{\bm{p}}}

\def\mSigma{{\bm{\Sigma}}}

\DeclareMathAlphabet{\mathsfit}{\encodingdefault}{\sfdefault}{m}{sl}
\SetMathAlphabet{\mathsfit}{bold}{\encodingdefault}{\sfdefault}{bx}{n}

\DeclarePairedDelimiter{\norm}{\lVert}{\rVert} 

\interspeechcameraready 

\title{Zero-Shot Mono-to-Binaural Speech Synthesis}

\author[affiliation={1}]{Alon}{Levkovitch}
\author[affiliation={2}]{Julian}{Salazar}
\author[affiliation={2}]{Soroosh}{Mariooryad}
\author[affiliation={2}]{RJ}{Skerry-Ryan} 
\author[affiliation={1}]{Nadav}{Bar} 
\author[affiliation={2}]{Bastiaan}{Kleijn}
\author[affiliation={1}]{Eliya}{Nachmani}

\affiliation{Google Research}
\affiliation{Google DeepMind}

\email{alevkovitch@google.com, eliyn@google.com}


\begin{document}

\maketitle

\begin{abstract}
   We present ZeroBAS, a neural method to synthesize binaural speech from monaural speech recordings and positional information without training on any binaural data. To our knowledge, this is the first published zero-shot neural approach to mono-to-binaural speech synthesis. Specifically, we show that a parameter-free geometric time warping and amplitude scaling based on source location suffices to get an initial binaural synthesis that can be refined by iteratively applying a pretrained denoising vocoder. Furthermore, we find this leads to generalization across room conditions, which we measure by introducing a new dataset, TUT Mono-to-Binaural, to evaluate state-of-the-art monaural-to-binaural synthesis methods on unseen conditions. Our zero-shot method is perceptually on-par with the performance of supervised methods on previous standard mono-to-binaural dataset, and even surpasses them on our out-of-distribution TUT Mono-to-Binaural dataset.

\end{abstract}

\section{Introduction}
\label{introduction}

Humans possess a remarkable ability to localize sound sources and perceive the surrounding environment through auditory cues alone. This sensory ability, known as \textit{spatial hearing}, plays a critical role in numerous everyday tasks, including identifying speakers in crowded conversations and navigating complex environments. Hence, emulating a coherent sense of space via listening devices like headphones becomes paramount to creating truly immersive artificial experiences. Due to the lack of multi-channel and positional data for most acoustic and room conditions, the robust and low/zero-resource synthesis of binaural audio from single-source, single-channel (mono) recordings is a crucial step towards advancing augmented reality (AR) and virtual reality (VR) technologies.

\begin{figure}
\includegraphics[width=0.47\textwidth]{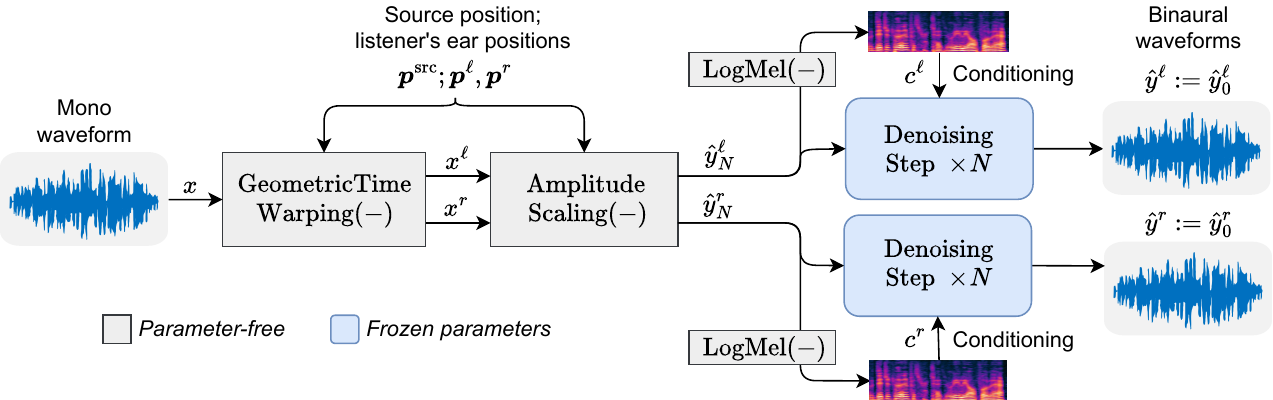}
\caption{Our proposed ZeroBAS method. Mono waveform is binauralized with geometric time warping conditional on the speaker's position, then the two channels' amplitudes are scaled. Each channel is then denoised 3 times a monaural denoising vocoder.}
\label{fig:overview}
\vspace{-0.5cm}
\end{figure}

Conventional mono-to-binaural synthesis techniques leverage a digital signal processing (DSP) framework. Within this framework, the head-related transfer function (HRTF), the room impulse response (RIR), and ambient noise are modeled as linear time-invariant (LTI) systems \cite{savioja1999creating, zotkin2004rendering, jianjun2015natural, zhang2017surround}. These DSP-based approaches are prevalent in commercial applications due to their established theoretical foundation and their ability to generate perceptually realistic audio experiences. However, real acoustic propagation, unlike the one modeled by LTI systems, has nonlinear wave effects. Recent advancements in the field have witnessed a paradigm shift towards employing machine learning methods via the paradigm of supervised learning \cite{richard21-warpnet, huang2022end, leng22-binauralgrad, lee23-nfs, liu23-dopplerbas, chen2023novel}. The task of synthesizing binaural audio from monophonic sources presents a significant challenge for supervised learning models. This difficulty stems from two primary limitations: (1) the scarcity of position-annotated binaural audio datasets, and (2) the inherent variability of real-world environments, characterized by diverse room acoustics and background noise conditions. Data collection for supervised learning necessitates specialized equipment, including tracking systems and binaural recording devices, which are both cost-prohibitive and often unavailable. Moreover, supervised models are susceptible to overfitting on the specific rooms, speaker characteristics, and languages in the training data, especially when the data is small (the standard dataset of \cite{richard21-warpnet} is only two hours). To address these limitations, we propose a novel zero-shot approach for monaural-to-binaural synthesis that is effective across a broader spectrum of recording scenarios by leveraging a monaural vocoder trained on tens of thousands of hours (\Cref{fig:overview}). Our contributions are:
\begin{itemize}
    \item The first zero-shot method for neural mono-to-binaural speech synthesis, leveraging geometric time warping, amplitude scaling, and a (monaural) denoising vocoder [WaveFit; \cite{koizumi22-wavefit}]. Notably, we achieve natural binaural speech generation that is perceptually on par (MOS, MUSHRA) with existing supervised methods despite never seeing binaural data.
    \item A novel dataset-building approach and dataset, TUT Mono-to-Binaural, derived from the location-annotated ambisonic recordings of speech events in the TUT Sound Events 2018 dataset \cite{adavanne2018sound}. When evaluated on this out-of-distribution data, past supervised methods degrade significantly while ZeroBAS continues performing well.
\end{itemize}

\section{Related Work}
\label{related-work}

DSP techniques approach the mono-to-binaural problem as a stack of acoustic components, each of which is an LTI system. Accurate wave-based simulation of RIRs is computationally expensive, and thus most real-time systems rely on simplified geometrical models \cite{valimaki2012fifty, savioja1999creating}. HRTFs need to be recorded inside an anechoic chamber in about 10k locations for good results \cite{li2020measurement}. DSP approaches treat these functions as a series of convolutions that are applied to the input signal. \cite{richard21-warpnet} proposed one of first uses of neural networks for mono-to-binaural synthesis, composing a neural time-warping module (WarpNet) and a temporal (hyper-)convolutional neural network (CNN) to learn a direct map between mono and binaural waveforms. BinauralGrad \cite{leng22-binauralgrad} was the first to use a denoising diffusion probabilistic model (DDPM). Since then, better incorporation of the inductive biases from DSP have led to more efficient neural systems. Neural Fourier Shift [NFS; \cite{lee23-nfs}] predicts delays and scaling from speaker locations and achieve close to state-of-the-art performance with a significantly smaller model. DopplerBAS \cite{liu23-dopplerbas} found that incorporating the Doppler effect into the conditioning features improved the phase loss of both the WarpNet and BinauralGrad systems. \cite{kitamura2023binaural} used a structured state space sequence (S4) model for the mono-to-binaural task. \cite{huang2022end} show that mono-to-binaural audio synthesis can be performed end to end with the use of audio codes. Motivated by the difficulty of collecting HRTF and RIR data, \cite{gebru21-implicit-hrtf} showed that an implicit HRTF can be learned by a temporal CNN. \cite{richard2022deep} and \cite{lee2022differentiable} showed that DNNs can be used to estimate RIR filters. \cite{luo2022learning} created a model for learning an implicit representation of an acoustic field. Furthermore, a different line of work uses visual conditioning for the generation of binaural audio \cite{chen2023everywhere,liang2024av,somayazulu2024self,yoshida2023binauralization,xu2024sounding}. 

\section{Approach}
\label{approach}
Our proposed zero-shot mono-to-binaural synthesis method utilizes a three-stage architecture. The first stage follows previous work and performs geometric time warping (GTW) to manipulate the input mono waveform into two channels based on the provided position information. Subsequently, our proposed amplitude scaling (AS) module adjusts the amplitude of the warped signal. Finally, an existing denoising vocoder iteratively refines the processed signal to generate the binaural output composed of two channels. Figure~\ref{fig:overview} provides a visual representation of this pipeline. Let $x$ denote the mono source signal. Its position at time $t$ is given by the 3D vector $\vp^{\text{src}}_t$. Let $\ell$ and $r$ correspond to the listener's left and right ear. Their positions at $t$ are given by 3D vectors $\vp^{\ell}_t, \vp^{r}_t$. The system first applies GTW to $x$ conditioned on $\vp^{\text{src}}_t, \vp^{\ell}_t$ and $\vp^{r}_t$. This warping gives left and right preprocessed channels, denoted by $x^{\ell}$ and $x^{r}$. Then, AS is employed jointly on $x^{\ell}$ and $x^{r}$, conditioning on the same data. This step aims to further enhance the spatial perception of the signal. The resulting intermediate left and right channels are denoted by $\hat{x}^{\ell}$ and $\hat{x}^{r}$, respectively. Finally, the denoising step sets its noisy inputs $\hat{y}^{\ell}_N, \hat{y}^{r}_N$ to be the outputs of the previous stage, $\hat{x}^{\ell}, \hat{x}^{r}$.  This replaces the typical Gaussian noise initialization used when training or sampling from denoising models. $\hat{y}^{\ell}_N, \hat{y}^{r}_N$ are fed separately into the same pretrained denoising vocoder, which treats each waveform as mono audio. The temporal sequences of conditioning vectors $\vc^l$, $\vc^r$ are obtained by extracting the log-mel features of $\hat{x}^{\ell}, \hat{x}^{r}$. A low noise level $k$ is also conditioned on, to reflect that we are emulating an input that is ``close'' to a true binaural sample. In the case of our denoising vocoder, WaveFit \cite{koizumi22-wavefit}, this noise level is given by a choice of conditioning timestep; specifically, the last timestep of the WaveFit training's denoising process. This sampling is repeated for $N$ iterations. 
In the Experiments section, we show that our approach produces a binaural speech rendering whose quality approximates the ground truth binaural audio. Note that our method does not take into account room effects nor the listener's head shape. Thus, we produce spatial audio which imputes both a generalized low RIR room (regularized by all the data the vocoder was trained on), and an implicit HRTF. 

\begin{algorithm}[t]
\begin{algorithmic}
\caption{ZeroBAS algorithm:}\label{alg:ZeroBAS}
\Require Denoising vocoder $\mathcal{V}_{\theta}$, iteration count $N$, low noise level $k$, mono waveform $x$, speaker position $\vp^{\text{src}}$, listener's ear locations $\vp^{\ell}, \vp^{r}$.
\State $x^\ell$, $x^r$ $ = \text{GeometricTimeWarping}(x, \vp^{\text{src}}, \vp^{\ell}, \vp^{r})$
\State $\hat{x}^{\ell}$, $\hat{x}^r$ $ = \text{AmplitudeScaling}(x^{\ell}, x^r, \vp^{\text{src}}, \vp^{\ell}, \vp^{r})$
\State $\vc^{\ell}, \vc^{r} = \text{LogMel}(\hat{x}^{\ell}),\text{LogMel}(\hat{x}^{r})$
\State $\hat{y}^{\ell}_N, \hat{y}^{r}_N := \hat{x}^{\ell}, \hat{x}^{r}$
\For{$i \gets N$ to $1$}
    \State$\hat{y}^{\ell}_{i-1}, \hat{y}^{r}_{i-1}  = \mathcal{V}_{\theta}(\hat{y}^{r}_{i}, \vc^{r}, k), \mathcal{V}_{\theta}(\hat{y}^{r}_{i}, \vc^{r}, k)$
\EndFor
      \State \Return $\hat{y}^{\ell}, \hat{y}^{r} := \hat{y}^{\ell}_0, \hat{y}^{r}_0$.
\end{algorithmic}
\end{algorithm}

\subsection{Geometric Time Warping (GTW)}

GTW aims to estimate a warpfield that separates the left and right binaural signals by applying the interaural time delay (ITD) based on the relative positions of the sound source and the listener's ears. \cite{richard21-warpnet} proposed GTW as a method to generate an initial estimate of the perceived signals. Let $S$ denote the signal's sample rate and $\nu_{\text{sound}}$
represent the speed of sound. The system employs basic GTW on the monaural signal $x$. This warping is achieved by computing a warpfield for both the left and right listening channels, denoted by $\rho^{\ell}(t), \rho^{r}(t)$. The values of this warpfield are computed using on the source and listener ear positions $\vp^{\text{src}}_t, \vp^{\ell}_t, \vp^{r}_t$:
\begin{align}
&\rho^{\ell/r}(t) := t - \frac{S}{\nu_{\text{sound}}}\;||\vp^{\text{src}}_t - \vp_t^{\ell/r}||_2
\end{align}
As this function takes non-integer values, the warped left and right signals $\hat{x}^{\ell}, \hat{x}^r$ can be defined with respect to the original indexing $t$ via linear interpolation.

\subsection{Amplitude Scaling (AS)}

 Human spatial perception of sound relies on various factors, including the ITD, the interaural level difference (ILD), and spectral cues due to HRTFs. While prior works \cite{wersenyi2010representations, baumgarte2003binaural} suggest that the ILD is mostly caused by scattering off of the head and is dominant in human spatial perception for sounds with high frequencies, we find that amplitude scaling based on the inverse square law has a positive effect on the perceived spatial accuracy of the processed signal. Our approach aims to leverage this amplitude manipulation to enhance the spatial realism of the generated binaural audio. Let $D$ be the Euclidean distance from the origin of the sound waves. Then by the inverse-square law, pressure drops at a $1/D^2$ ratio \cite{zahorik2005auditory}. In the case of microphones, pressure manifests as amplitude. Acknowledging that the left-right microphone distance of the KEMAR mannequin used in datasets like \cite{richard21-warpnet} is only an approximation of human heads, we define:
\begin{align}
    D_t^{\ell} &= \norm {\vp^{\text{src}} - \vp^{\ell}_t}_2, \quad \quad D_t^{r} = \norm {\vp^{\text{src}} - \vp^{r}_t}_2.
\end{align}

Then, at each time step we scale down the magnitude of the side furthest from the source, using the ratio of the closer side's distance versus the further side's distance:
\begin{align}
    &\hat{x}_t^{\ell} := \min(1, (D_t^{r}/D_t^{\ell})^2) \cdot x_t^{\ell},\\ &\hat{x}_t^{r} := \min(1, (D_t^{\ell}/D_t^{r})^2) \cdot x_t^{r}.
\end{align}

\subsection{Denoising Vocoder}

GTW and AS are simple, parameter-free operations that only roughly approximate binaural audio; using the warped and scaled speech signals $\hat{x}^{\ell}, \hat{x}^{r}$ as-is results in acoustic artifacts and inconsistencies. Hence, there is a need for further refinement to generate natural-sounding binaural audio. To this end, we propose that a sufficiently well-trained denoising vocoder could be used on each signal \textit{independently}. We use a WaveFit neural vocoder \cite{koizumi22-wavefit} as our denoising vocoder model. It is a fixed-point iteration vocoder combined with the discriminator of generative adversarial networks, specifically MelGAN's \cite{kumar19-melgan}, to learn a sampling-free iterable map that can generate natural speech from a degraded input speech signal. As a vocoder, it takes log-mel spectrogram features and noise as input and produces clean waveform output. In WaveFit's notation, we perform the iterated application of
\begin{align}
\label{wavefit}
    \hat{y}_{i-1} := \mathcal{V}_{\theta}(\hat{y}_i, \vc, k) := \mathcal{G}(\hat{y}_i - \mathcal{F}_{\theta}(\hat{y}_i, \vc, k), \vc),
\end{align}
where $\vc$ is the spectrogram to convert and $\hat{y}_{i-1}$ is a candidate waveform refined from $\hat{y}_i$. $\mathcal{G}$ is a parameter-free gain adjustment operator and $\mathcal{F}_\theta$ is the WaveGrad architecture \cite{chen21-wavegrad} trained for reconstruction under a discriminator. At training time, the starting noise is given by $\hat{y}_K \sim \mathcal{N}(0, \mSigma_\vc)$ where $\mSigma_\vc$ is a covariance matrix initialized as in SpecGrad \cite{koizumi22-specgrad} to capture the spectral envelope of $\vc$; both $k, i$ iterate over $K, \dotsc, 1$. Then, at inference time, we express our ``approximation'' hypothesis by iterating at the noise level of WaveFit's final denoising step ($k = 1$). We then iteratively denoise $\hat{y}^{\ell}_N, \hat{y}^{r}_N := \hat{x}^{\ell}, \hat{x}^{r}$, conditioning on their initial log-mel spectrograms and the fixed low noise level for steps $i = N, \dotsc, 1$.

\section{Experiments}

\subsection{Data and Models}

For our experiments we use two datasets. The first is the Binaural Speech dataset (BSD) released by \cite{richard21-warpnet}. The dataset contains paired mono and binaural audio with tracking information, jointly collected in a non-anechoic room; see \cite{richard21-warpnet} for more details. The second dataset is an adapted version of TUT Sound Events 2018 \cite{adavanne2018sound} which we name "TUT Mono-to-Binaural" (TMB). It contains 1,174 recordings, each about 2 seconds long. Overall, there are 2.15 hours of recordings in the dataset. The spoken language is French, speakers are recorded in a studio, and each recording is played in a single location. Using this dataset ensures a zero-shot evaluation for all of the methods tested in this paper, as none were trained on this data. For our DSP baseline, we use the open-source Resonance Audio package. The WaveFit vocoder is described in \cite{koizumi22-wavefit}. The pretrained weights we use were trained on the 60k-hour LibriLight audiobook dataset \cite{kahn2020libri} as described in \cite{koizumi22-wavefit}.

\subsection{TUT Mono-to-Binaural: Dataset Construction}

Our purpose in creating and using the TMB dataset is threefold: (a) demonstrate a new approach for creating mono-to-binaural synthesis datasets due to their scarcity, (b) evaluate the ability of different methods to generalize to different rooms and acoustic environments, and (c) evaluate the ability of different methods to generalize to different speakers. The TUT Sound Events 2018 is build for sound event localization and is composed of ambisonic recordings from the DCASE 2016, Task 2 dataset. In TUT Sound Events 2018, mono recordings were played back using a loudspeaker at distances ranging from $1$-$10$ meters and captured by an ambisonic microphone. Sound event locations are given using azimuth, elevation and distance, with each sound event having a single location. Starting from this data, we apply several processing steps: Frist, speaker location information provided in azimuth, elevation, and distance were converted into a Cartesian coordinate system $(x, y, z)$. Next, ground-truth metadata was leveraged to cut out speech segments from the recordings using their provided timestamps. To generate the binaural ground truth for our evaluation, ambisonic recordings are converted to binaural audio using OmniTone, a well-established DSP ambisonic decoder with a binaural renderer. Finally, the corresponding original monaural recordings are obtained from the DCASE 2016, Task 2 dataset. Note that unlike the BSD, these mono recordings are recorded separately from their ambisonic re-recordings and binaural renderings.

\subsection{Evaluations}

For objective evaluations, we use metrics found in prior work. \textbf{Wave (W) $\pmb{\ell_2}$}: mean squared error (MSE) between the ground truth and synthesized per-channel waveforms multiplied by $10^3$. \textbf{Amplitude (A) $\pmb{\ell_2}$}: MSE between the amplitude STFTs of the ground truth and synthesized audio. \textbf{Phase (P) $\pmb{\ell_2}$}: MSE between the left - right phase angle of the ground truth and synthesized audio. Phase is computed from the STFT. \textbf{MRSTFT $\mathcal{L}_{\text{S}}$} is the multi-resolution spectral loss. For subjective evaluations, we perform \textbf{MOS} and \textbf{MUSHRA} evaluations. For MOS, we collect mean opinion scores towards axes of naturalness. 

For every experiment, we use $50$ random samples from each method. Every example is rated $5$ times by different raters, with each experiment participated in by at least $30$ raters. In the MUSHRA evaluation, 

we used $50$ random samples from each method. Following the MUSHRA protocol, we discard raters who gave >15\% of hidden references a score below 90. We used the model and code releases of WarpNet, BinauralGrad, and NFS to synthesize audio for subjective evaluations of these systems.

\addtolength{\tabcolsep}{-2.5pt}
\begin{table}[t]
\caption{Objective and subjective evaluations on BSD}
\label{tbl:objective_orig}
\begin{small}
\begin{tabular}{@{}llccccc@{}}
\toprule
&Model & W $\ell_2$ $\downarrow$ & A $\ell_2$ $\downarrow$ & P $\ell_2$ $\downarrow$ & $\mathcal{L}_{\text{S}}$ $\downarrow$ & MOS $\uparrow$\\
\midrule
 {\fontsize{7pt}{0}\selectfont Zero-} & DSP& 0.812 & \textbf{0.052} & 1.572 & 1.91 & 3.84$\pm$0.19\\
 {\fontsize{7pt}{0}\selectfont Shot} & ZeroBAS& \textbf{0.440} & 0.053 & \textbf{1.508}& 1.91 & \textbf{4.07$\pm$0.17}\\
\midrule
 {\fontsize{7pt}{0}\selectfont Sup-} &WarpNet & 0.179 & 0.037 & 0.968 & 1.52 &3.86$\pm$0.16 \\
 {\fontsize{7pt}{0}\selectfont ervi-}& BGrad  & \textbf{0.128} & \textbf{0.030} & \textbf{0.837} & \textbf{1.25} & \textbf{4.01$\pm$0.14}\\
 {\fontsize{7pt}{0}\selectfont sed} &NFS & 0.172 & 0.035 & 0.999 & 1.29 &3.99$\pm$0.15 \\
\midrule 
& GT & - & - & - & - & 4.30$\pm$0.12 \\
\bottomrule
\end{tabular}
\end{small}
\end{table}

\begin{figure}[t]
\vspace{-0.2cm}
    \begin{subfigure}{0.22\textwidth}
        \includegraphics[width=\textwidth]{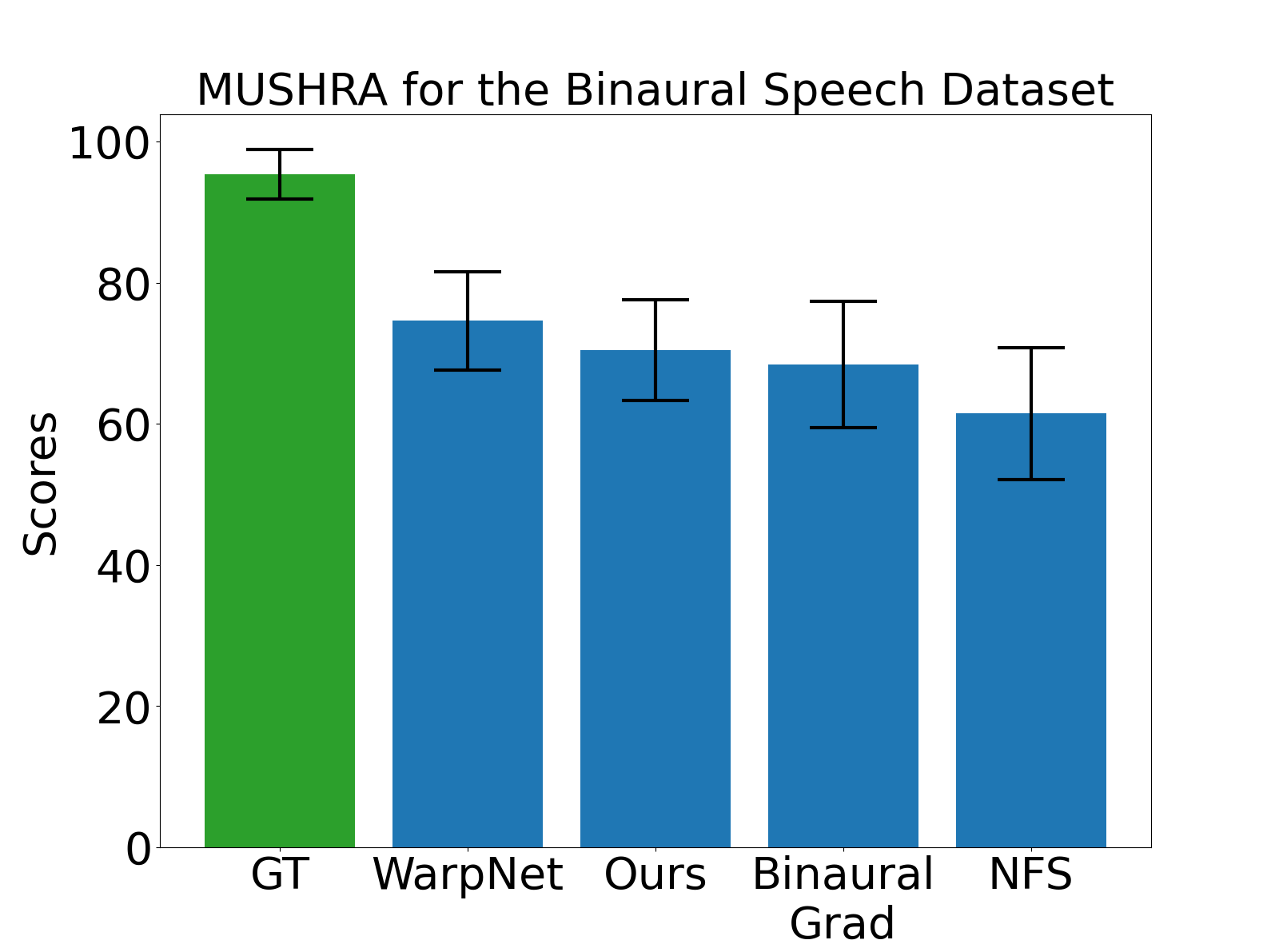}
        \caption{}
        \label{fig:mushra_orig}
    \end{subfigure}
    \hfill
    \begin{subfigure}{0.22\textwidth}
        \includegraphics[width=\textwidth]{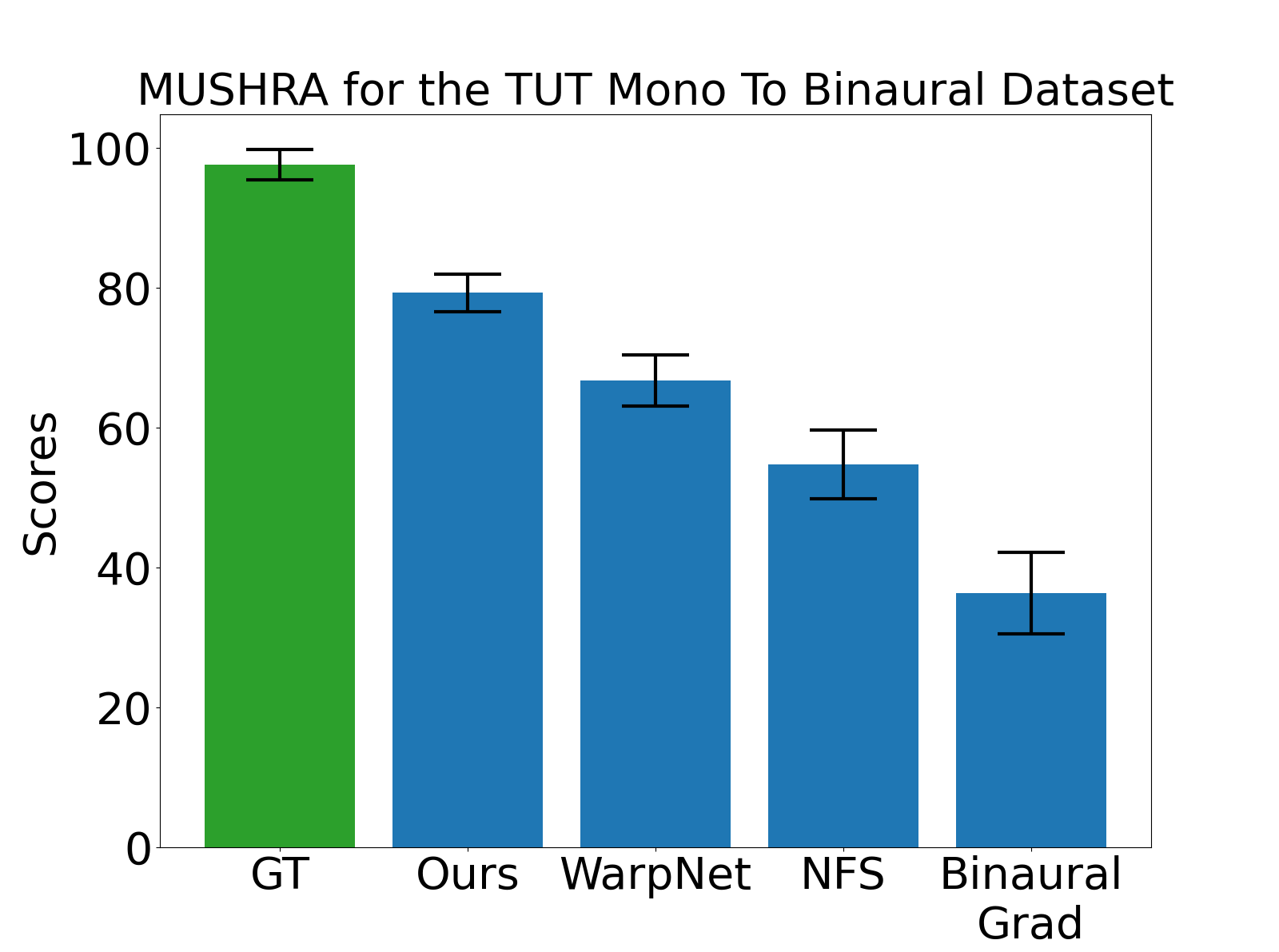}
        \caption{}
        \label{fig:mushra_tut}
    \end{subfigure}
    
    \caption{MUSHRA results for (a) the BSD and (b) the TMB.}
    \label{fig:mushra_subplots}
    \vspace{-0.7cm}
\end{figure}

\subsection{Binaural Speech Dataset Results}

In \Cref{tbl:objective_orig}, we observe that ZeroBAS achieves significant objective improvements over the DSP baseline, despite not modeling additional interactions between the two generated channel streams or the RIR and HRTF. Furthermore, the performance of the ZeroBAS method approaches that of the supervised methods, even though ZeroBAS has not been trained on the BSD. 
Note that ZeroBAS inherently cannot model certain imperceptible environment-specific artifacts, like high-frequency recording equipment noise. Supervised methods may capture these and get superficial improvements on objectives like phase error, whereas vocoders may explicitly ignore them.

In fact, subjective evaluation results in \Cref{tbl:objective_orig} show that ZeroBAS sounds slightly more natural to human raters than the supervised methods while being on par in comparative evaluations. 
MUSHRA results (\Cref{fig:mushra_subplots}) no statistically significant preference for any of the methods WarpNet, BinauralGrad, NFS or ZeroBAS. The ZeroBAS system leverages a WaveFit model which ensures the generated audio exhibits minimal artifacts and noise compared to binaural recordings, leading to improved perceptual quality for human listeners. Despite worse objective metrics, our human evaluations suggest ZeroBAS achieves spatial fidelity and quality on par, if not better than supervised methods. Samples can be heard in our demo page: \url{https://alonlevko.github.io/zero-bas/}. 

\subsection{TUT Mono-to-Binaural Results}

\begin{table}[t]
\caption{Objective and subjective evaluations on TMB.}
\label{tbl:objective_tut}
\begin{small}
\begin{tabular}{@{}llccccc@{}}
\toprule
& Model & W $\ell_2$ $\downarrow$ & A $\ell_2$ $\downarrow$ & P $\ell_2$ $\downarrow$ & $\mathcal{L}_{\text{S}}$ $\downarrow$ & MOS $\uparrow$ \\
\midrule
{\fontsize{7pt}{0}\selectfont Zero-} & DSP    & 1.134 & 0.075 & 1.572 & 2.93 & 3.09$\pm$0.28 \\
{\fontsize{7pt}{0}\selectfont Shot} & ZeroBAS          & \textbf{0.293} & \textbf{0.045} & 1.572 &  \textbf{2.93} & \textbf{3.98$\pm$0.15} \\
\midrule
{\fontsize{7pt}{0}\selectfont Sup-} & WarpNet       & 2.909 & 0.099 & 1.571 & 6.66 & 3.60$\pm$0.26\\
{\fontsize{7pt}{0}\selectfont ervi-} & BGrad  & 3.228 & 0.218 & 1.571 &  5.40 & 3.27$\pm$0.32 \\
{\fontsize{7pt}{0}\selectfont sed} & NFS           & \textbf{1.574} & \textbf{0.085} & 1.571 & \textbf{3.06} & \textbf{3.79$\pm$0.23}\\
\midrule
& GT & - & - & - & - & 4.08$\pm$0.11 \\
\bottomrule
\end{tabular}
\end{small}
\vspace{-0.4cm}
\end{table}

Although the zero-shot method underperforms supervised methods in the subjective evaluation on BSD, we argue that the supervised methods are sensitive to the room and recording conditions of BSD. To demonstrate this, we evaluated all methods on our newly constructed TMB. \Cref{tbl:objective_tut} demonstrates that our zero-shot method, ZeroBAS, significantly outperforms all supervised methods on TMB. 
Both ZeroBAS and the supervised methods struggle to capture accurate phase information, as evidenced by P $\ell_2$. 

The subjective evaluation results presented in \Cref{tbl:objective_tut} further demonstrate that ZeroBAS exhibits superior performance in terms of perceived naturalness compared to the supervised methods WarpNet, BinauralGrad, and NFS. As evidenced by the MOS, ZeroBAS surpasses these methods by notable margins. 
Considering the confidence intervals, these results indicate that human listeners on TMB perceive ZeroBAS as more natural-sounding than the supervised methods, with its score approaching that of the ground truth recordings. Furthermore, MUSHRA evaluations reveal a statistically significant preference for the proposed ZeroBAS method compared to supervised approaches. This suggests that human listeners perceive the spatial quality of binaural signals generated by ZeroBAS to best align with the reference. Evaluation of existing supervised learning methods on TMB revealed several limitations. BinauralGrad produced outputs with substantial Gaussian noise, hindering the diffusion process's convergence to clean signals for out-of-distribution samples. WarpNet and NFS exhibited two key failure modes: (a) Inability to retain speaker voice characteristics in the binaural output, leading to fidelity degradation, and (b) incorrect spatialization, manifesting as generated binaural speech with unrealistic distance cues or spatial artifacts when beyond the training range. These failures are further illustrated here: \url{https://alonlevko.github.io/zero-bas/}.

\section{Ablation Analysis}

\begin{table}[t]
\centering
\caption{Ablation of our ZeroBAS method on BSD}
\label{tbl:ablations}
\begin{small}
\begin{tabular}{@{}lcccc@{}}
\toprule
Model & W $\ell_2$ $\downarrow$ & A $\ell_2$ $\downarrow$ & P $\ell_2$ $\downarrow$ & MOS $\uparrow$ \\
\midrule
ZeroBAS & 0.440 & 0.053 & 1.508 & 4.07$\pm$0.17  \\
\midrule
w/o AS & 0.802 & 0.059 & \textbf{1.539} & 2.93$\pm$0.16\\
w/o GTW & 0.627 & 0.053 & 1.569 & 3.64$\pm$0.15\\
w/o AS, GTW & 0.816 & 0.051 & 1.567 & \textbf{4.13$\pm$0.18}\\
w/o WaveFit & \textbf{0.539} & \textbf{0.044} & 1.572 & 3.52$\pm$0.16 \\
\midrule
Original Decode & 0.495 & \textbf{0.065} & 1.534 & 2.50$\pm$0.16\\
Swap Order & \textbf{0.474} & 0.072 & \textbf{1.277} & \textbf{3.85$\pm$0.19}\\
\midrule
1 iteration & 0.459  & 0.069 & \textbf{1.393} &3.62$\pm$0.20 \\
2 iterations & 0.450 & 0.061 & 1.492 &  3.83$\pm$0.24\\
4 iterations & \textbf{0.445 }& \textbf{0.053} & 1.502 & 3.94$\pm$0.18\\
5 iterations & 0.449 & \textbf{0.053} & 1.494 &  \textbf{4.05$\pm$0.15} \\
\bottomrule
\end{tabular}
\end{small}
\vspace{-0.4cm}
\end{table}

The significance of each core component is evaluated through ablation studies (Table~\ref{tbl:ablations}). First, AS is critical for ZeroBAS performance. Its absence leads to substantial degradation in both MOS and W $\ell_2$. AS creates a crucial perceptual difference. GTW is the second most important component. Without GTW, left-right channel time differences become misaligned, resulting in increased W $\ell_2$ error and decreased MOS. The WaveFit model, when removed in isolation, has a minimal impact on objective metrics but a significant negative impact on MOS which highlights it's importance. Removing both AS and GTW leads to improved MOS, albeit resulting in a monaural waveform played identically in both channels. In addition we tested the effects of modifications within WaveFit. Decoding for five iterations and initializing with Gaussian noise (Original Decode), as in the original WaveFit implementation, resulted in poor audio quality. This is because the two channels remain independent, and playing them as a binaural recording produces an unaligned and noisy output. Furthermore, applying WaveFit to the monaural input first (Swap Order), followed by AS and GTW, yielded improved performance in terms of P $\ell_2$ but compromised MOS and A $\ell_2$. Finally, increasing the number of WaveFit iterations until 3 improves the objective metrics W $\ell_2$, A $\ell_2$ and P $\ell_2$ and improves MOS. After 3 iterations, quality is constant.

\section{Conclusion}
In this work, we presented a room-agnostic zero-shot method for binaural speech synthesis from monaural audio. Our results demonstrate that the method achieves perceptual performance comparable to supervised approaches on their in-distribution datasets. Furthermore, we introduce a novel dataset designed to evaluate the generalization capabilities of monaural-to-binaural synthesis methods for out-of-distribution scenarios. On this dataset, ZeroBAS exhibits superior performance compared to supervised methods, highlighting its potential for real-world applications with diverse acoustic environments.

\clearpage
\bibliographystyle{IEEEtran}
\bibliography{mybib}

\clearpage

\end{document}